\documentstyle[12pt,pictex,epsfig]{article}
\newcommand{\ra}{\rightarrow}

\newcommand{\be}{\begin{equation}}
\newcommand{\ee}{\end{equation}}
\newcommand{\bea}{\begin{eqnarray}}
\newcommand{\eea}{\end{eqnarray}}

\newcommand{\ov}{\overline}

\begin{document}
\thispagestyle{empty}   
\begin{flushright}
August, 1996
\end{flushright}
\begin{flushleft}
Oslo TP 3-96
\end{flushleft}
\vspace{0.4cm}

\vspace{0.5cm}

\begin{center}
  \begin{Large}

  \begin{bf}
The self-penguin contribution to $K \rightarrow 2 \pi$
  \end{bf}
  \end{Large}
\end{center}

  \vspace{0.5cm}
  \begin{Large}
\begin{center}     
A. E. Bergan\footnote{e-mail : elias.bergan@fys.uio.no}
 and J. O. Eeg\footnote{e-mail :j.o.eeg@fys.uio.no}

\end{center}
\end{Large}

\begin{Large}
\begin{center}
{\small  Department of Physics, University of Oslo, \\

P.O.Box 1048, N-0316 Oslo, Norway \\}
\end{center}
\end{Large}
\vspace{0.6cm}
\begin{center}
  {\bf Abstract}
\end{center}
\begin{quotation}
\noindent
We consider the  contribution to $K \rightarrow 2 \pi$ decays 
from the non-diagonal $s \ra d$ quark transition amplitude. 
First, we calculate the most important part of the $s \rightarrow d$
transition, the so-called self-penguin amplitude $\sim G_F \alpha_s$,
including the heavy top-quark case.
Second, we calculate the matrix element of the $s \rightarrow d$
transition   for the physical
 $K \rightarrow 2 \pi$ process. This part of the
analysis is performed within the Chiral Quark Model
 where 
quarks are coupled to the
pseudoscalar mesons.

The CP-conserving self-penguin 
 contribution to $K \rightarrow 2\pi$ is found to be negligible. 
The  obtained contribution to $\epsilon'/\epsilon$ is sensitive to the 
values of the quark condensate $<\bar{q} q>$ and the constituent quark
mass $M$. For reasonable values of these quantities we find that the
 self-penguin
contribution to $\epsilon'/\epsilon$
is  10-15 $\%$ of the gluonic penguin
 contribution and has the same sign. 
Given the large cancellation between gluonic and electroweak penguin
contributions,
this means that our contribution is of the same order of magnitude
as $\epsilon'/\epsilon$ itself.

\end{quotation}

\newpage

\begin{Large}

1. Introduction

\end{Large}

\vspace{0.3cm}
 In general,
non-leptonic $\Delta S = 1$ decays are  described by an effective 
Lagrangian at quark level \cite{GaL,AlMa,NLept}
\begin{equation}
{\cal{L}}(\Delta S =1)  =  \sum_{i} C_i Q_i  \; \;  .
\label{efflag}
\end{equation}
 In such effective Lagrangians, the heavy
mass scales are integrated out and their effects are contained in the Wilson
coefficients $C_i$. The  quark operators $Q_i$ involve the
three light quarks $q = u,d,s$. The coefficients will in general
include short distance QCD effects calculated by means of
perturbation theory and the
renormalization group equations (RGE).
In our notation the Wilson
coefficients $C_i$ also contain  Fermis coupling constant $G_F$
 and  the relevant
Kobayashi-Maskawa factors $\lambda_q = V_{sq} V^*_{dq}$
for CP-conserving ($q=u$) and CP-violating ($q=t$)
$\Delta S = 1$ transitions respectively.

 Within this
standard procedure, one usually omits operators containing
 $(i \gamma \cdot D - m_q)$,
 by appealing to the QCD equations of motion for quark fields
 \cite{AlMa,CoDDPoS,OP}:
\begin{equation}
 (i \gamma \cdot D - m_q) \rightarrow 0 \; ,
\label{onshell}
\end{equation}
where $D_{\mu}$ is the covariant derivative containing the gluon
 field, and $m_q$ the appropriate quark mass.
This procedure corresponds to going on-shell with external
quarks in quark operators. Certainly, quarks are not 
on-shell in hadrons, especially not in the  (would-be
Goldstone) mesons $\pi$  and $K$.

In a series of papers \cite{Do,ShaP,SPj,offsh}, it has been pointed out that
the effect of  such operators are not zero in general, and that they
in particular will  
contribute to processes like $K \rightarrow 2 \pi$ ,
$K \rightarrow 2 \gamma$ , and $B \rightarrow 2 \gamma$.
In order to be consistent, such genuinely off-shell effects
should go to zero in the limit when
some typical bound state parameter goes to zero, which is indeed found to be
the case. In our case this parameter is the constituent quark mass of the 
light $u,d,s$ quarks. Numerically, 
the non-zero off-shell effects  are non-negligible in some cases.

In this paper we will study
a  quantity which by construction  vanishes on the $s$- and $d$- quark 
mass-shells
after regularization, namely the non-diagonal $s\rightarrow d$
self-energy transition $\Sigma_{ds}$ \cite{Do,ShaP,SPj,GPP}.
 For off-shell quarks, however, this renormalized non-diagonal
self-energy is still non-zero and could a priori be relevant for
 physical amplitudes, e.g $K \rightarrow 2\pi$.
The unrenormalized $s\rightarrow d$ transition
corresponds to an effective Lagrangian
\begin{equation}
 {\cal{L}}_{ds}^U = \bar{d}(a i \gamma \cdot D L + b L + c R) s \; ,
\label{LdsU}
\end{equation}
where $a,b,c$ are divergent quantities given by the loop
integrations, and
$L, R$ are the left- and right-handed projectors in Dirac space.
In the Standard Model Lagrangian there are no direct $s \rightarrow d$
transitions. Consequently, one demands that the renormalized 
self-energy  vanishes on the respective mass shells of the $s$- and $d$-quarks,
by adding the necessary counterterms.
It should be emphasized that this defines the physical $s$ and $d$ quarks 
in the presence of weak interactions.
The  renormalized self-energy 
 corresponds to an effective Lagrangian of the form \cite{EPSG}
\begin{equation}
{\cal{L}}^R_{ds} = - A \, \bar{d} (i \gamma \cdot D - m_d)
  (i \gamma \cdot D R + M_R R + M_L L) 
(i\gamma \cdot D - m_s) s \; .
\label{LdsR}
\end{equation}
 In momentum space,
$A$ is a finite, slowly varying (logarithmic)
function of the $W$-boson and heavy quark masses and the external
 $s$- and $d$- quark
 momentum ($k$) squared.
To be explicit, $A$ is obtained from an expansion of  $a$ in (\ref{LdsU}):
 $a = a_0 + k^2 A + \cdots$. The quantity $a_0$, which is divergent and
independent of $k^2$, is removed by the counterterms.
 The quantities $M_{L,R}$ depend on the  masses
of the $s$- and $d$-quarks.

In the limit $m_{s,d} \ra 0$, which we will work,  the Lagrangian is
 \begin{equation}
{\cal{L}}^R_{ds}(m_{s,d} \ra 0) = - A \, \bar{d} (i \gamma \cdot D )^3
  L s \; .
\label{LdsR0}
\end{equation}

In the pure electroweak case, there is a strong suppression of the 
CP-conserving amplitude owing to the GIM-mechanism \cite{GIM}, and one 
finds \cite{GaL} that $A$ is of order $G_F m_c^2/M_W^2$.
 In the CP-violating case this strong GIM suppression
is relaxed due to  a top-quark with a mass of the
same order as the $W$-boson.  Even more important, when perturbative
 QCD to lowest order is added,
one obtains an unsuppressed charm-quark contribution 
 $\sim G_F \alpha_s (log \, m_c)^2$ \cite{ShaP,SPj}.
 A Feynman diagram for this contribution is shown in Fig.~1  and is named 
{\em self-penguin} because the gluon from a one loop penguin diagram is
reabsorbed by the external $s$- or $d$-quarks.

 If one applies the perturbative QCD 
 equations of motion as in (\ref{onshell}), we observe that
 ${\cal{L}}^R_{ds} \rightarrow~0$. However, we should emphasize that
 ${\cal{L}}_{ds}^R$ is obtained within
perturbation theory. 
Therefore, it is not allowed to apply (\ref{onshell})  for  quarks
strongly bound in $\pi, K$. 
An analogue within QED would be to replace (\ref{onshell}) by
\bea
 (\gamma \cdot D - m) \rightarrow  e \gamma  \cdot A^C \; ,  
\label{offshell}
\eea    
where the perturbative photons are contained in the covariant derivative, and
$A^C_\mu$ represents the binding Coulomb forces\cite{sak}. Also, in QED the
renormalized  electron self-energy
is zero on the electron mass-shell, but still it gives an important
contribution to Lamb-shift. As a gedanken experiment, if the $s$- and 
$d$-quarks were bound by electromagnetic forces, the contribution to
$(\bar{d} s) \ra 2 \gamma$ from (\ref{LdsR0}) would be proportional
 to the binding energy of the system, that is, of the order
$\alpha_{em}/r_B$, where $r_B$ is the Bohr radius. 
Being  small in QED, one expects  
 off-shell effects to be bigger in strong interactions where bound state
effects are more important. Thus,
physical effects for $K \rightarrow 2\pi$ decays from ${\cal{L}}_{ds}^R$ 
could be obtained, and  one
should explore possible consequences for the $\Delta I = 1/2$ rule
and for $\epsilon'/\epsilon$.


\begin{figure}
\begin{center}
\mbox{\epsfig{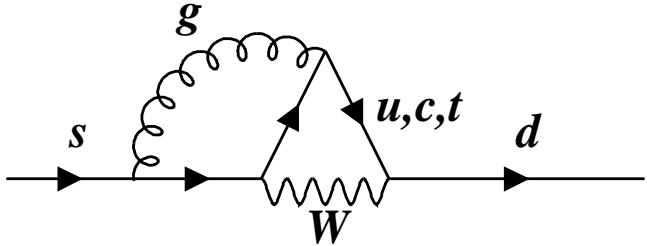}}
\end{center}
\label{sdselfenergy}
\caption{The self-penguin diagram. There is also a counterpart where the
gluon is absorbed by the $d$-quark.}
\end{figure}

\vspace{0,1cm}

To calculate the matrix elements of quark operators between physical
had\-ronic states is in general a difficult  task, and one  normally
 uses  various models or assumptions.
In this paper we will
 use the Chiral Quark Model ($\chi QM$),  advocated by many
authors \cite{ChiQM}.
This model consists of the ordinary QCD Lagrangian and 
 an extra term which induces  meson-quark couplings
that makes it possible to calculate matrix elements of quark
operators in terms of quark loop diagrams.
The model is thought to be applicable for quark momenta
 below some scale of the order $\Lambda_{\chi}$,  the  chiral symmetry
 breaking scale. 
We find this model very interesting because it reproduces the
 chiral Lagrangian terms for strong interactions to good  accuracy to order
$p^4$. Furthermore, it has been shown to be useful in calculating
weak hadronic matrix elements \cite{PdeR,BEF}.

\vspace{0.5cm}

\begin{Large}

2. The $s \rightarrow d$ self-penguin transition

\end{Large}

\vspace{0.2cm}

The renormalized non-diagonal  $s \ra d$  off-diagonal
 self-energy has the following form in agreement with (\ref{LdsR0})
\bea
{\cal M}(s \ra d) \; = \; 
-i  \bar{d} \, \Sigma_{ds} \, s  \, =
 \, - i k^2 \, \bar{d} \, \gamma \cdot k L s  A(k) \; \, ,
\label{Sigma}
\eea 
where $k$ is the four momentum of the external
$s$- and $d$- quark. In the pure electroweak (EW) case we obtain
for one flavour $q=u,c,t$:
\be
A(k)_{EW} \; = \; 2 \lambda_q \frac{g_W^2}{16\pi^2}
 (1 + \frac{m^2}{2M_W^2}) \int_0^1 dx \frac{x}{k^2}
\ln \left[ 1 - \frac{x(1-x)k^2}{\ov{M_x^2}} \right] \; \, ,
\label{ASig}
\ee
where $\ov{M_x^2} = x M_W^2 + (1-x)m^2$, $M_W$ is the $W$-boson mass and 
$m = m_q$ is the mass of the quark running in the loop ($q = u,c,t$),
$g_W$ is the $W$-coupling, and $\lambda_q$ is the appropriate KM factor.
 The term $\sim  m^2/(2M_W^2)$ is due to  unphysical Higgs
exchange in the Feynman gauge.
 For small $k^2$ we obtain
\be
A_{EW}  \; = \; \frac{G_F}{\sqrt{2}} \frac{1}{2\pi^2}  
\left[ \lambda_u (r_u -r_c) -  \lambda_t (r_c - r_t) \right] \; ,
\label{ewamp}
\ee
where the quantities $r_q$ are given by
\be
r_q  \; = \; - \,
(1 + \frac{m^2}{2M_W^2}) \int_0^1 dx \, x^2 (1-x)
\frac{M_W^2}{\ov{M_x^2}} \; ,
\label{ASig1}
\ee
for $q=u,c,t$.
As already mentioned, the pure electroweak CP-conserving 
$s \rightarrow d$ transition amplitude is very suppressed due to the
 GIM mechanism. A priori one expects a relaxed GIM mechanism for
 the CP-violating case involving the heavy top quark. The top quark
contribution is, in Feynman gauge, dominated by the unphysical Higgs
contribution. As a curiocity, we find a ``super-GIM'' mechanism:
It turns out that the top-quark contribution exactly cancels the
charm contribution ($(r_c - r_t) \ra 0$) in the limit
 $m_c \rightarrow 0$ and 
$m_t \rightarrow \infty$. Therefore the GIM-cancellation is somewhat
stronger than expected in the pure electroweak CP-violating case.
Numerically, we find $(r_c - r_t) =  -0.167 - (-0.124) = -0.043$
for $m_t = 180$ GeV.

As shown originally by Shabalin \cite{ShaP}, 
 the self-penguin diagram has
only a weak, logarithmic GIM suppression  to order $\alpha_s G_F$.
To obtain the two loop self-penguin we need  the penguin
 ($s \rightarrow d + gluon$) loop for arbitrary external momenta. By direct 
calculation this is divergent, and before we insert it into the next
 loop to get the self-penguin, it has to be properly 
regularized. For this purpose we consider the  Ward-identity  
\be
q_{\nu} \Gamma^{\nu}_{ds} =   \Sigma_{ds}(k + q) - \Sigma_{ds}(k) \; ,
\label{wardid}
\ee  
in order to determine the necessary counterterms.
In our case $\Gamma^{\nu}_{ds}$ corresponds to the penguin one loop
(the strong coupling $g_s$ not included).
Writing
\be
 \Gamma^{\nu}_{ds}(k,q) =   T^{\sigma \nu} \gamma_{\sigma} L \; ,
\label{Gamma}
\ee
the most general tensor  compatible with the Ward identity (\ref{wardid})
can be written
\bea
  T^{\sigma \nu} \; = \; 
\left[(k+q)^2 A(k+q) - k^2 A(k) \right] \frac{k^\sigma q^\nu}{q^2}
 + (k+q)^2 A(k+q) \, g^{\sigma \nu}     \nonumber \\
+ \, B_0 [q^\sigma q^\nu - q^2 g^{\sigma \nu}] \,
+ \, B_1 [q^\sigma k^\nu - q \cdot k g^{\sigma \nu}] \,  \nonumber \\
+ B_2 [k^\sigma k^\nu - \frac{q \cdot k}{q^2} k^\sigma q^\nu] \, 
+ B_\epsilon  i \epsilon^{\sigma \nu \alpha \beta} k_\alpha q_\beta
\; \,  .
\label{Tsinu}
\eea
Comparing this with the raw result for $\Gamma^{\nu}_{ds}$, we find the
regularized expressions:
\bea
B_j  =  2 g_W^2  \int_0^1 dx \int_0^{1-x}  dy f_j 
\left[\ov{M_y^2} -x(1-x)q^2 -y(1-y)k^2 + 2xy q \cdot k \right]^{-1} .
\label{Bfunc}
\eea
 For $W$- exchange only, the $f_j$'s
are given by $f_0 = 2x(1-x); f_1 = y(2x-1); f_2 = -2y^2$ and
$f_\epsilon = -y$, respectively. There are similar expressions for the
unphysical Higgs  case. 
We obtain the following expression for the renormalized
self-penguin contribution: 
\be
{\cal M}(s \ra d, g_s^2) \, = \,  -2i g_s^2 g_W^2 \ov{d} t^a t^a 
\int \frac{d^4q}{(2 \pi)^4} \frac{\gamma_{\nu}}{q^2}
  S(k+q) \Gamma^{\nu}_{ds}(k,q) \, s   \; ,
\label{SP1}
\ee
 where $t^a$ is a colour matrix.
To obtain the result compatible with (\ref{LdsR0}),  we expand
the $d$- (or $s$-) quark propagator $S(k+q)$ and
 $\Gamma^\nu_{ds}$ in powers of the external momentum $k$
and keep in total the order $(k)^3$ only. Then we obtain (for $q=c,t$):
\be
{\cal M}(s \ra d, g_s^2) \, = \, -i \frac{g_s^2 g_W^2}{3 M_W^2} 
 \lambda_q  k^2 k^{\mu}
 \ov{d}t^a t^a \gamma_{\mu} L s \left(\frac{i}{16 \pi^2}\right)^2  F_q \; ,
\label{SP2}
\ee 
where the dimensionless quantity:
\bea
 F_q & = & 6 M_W^2 \int_0^1 dx \int_0^{1-x} dy \left\{ 
 \frac{[(-1/2+2 x)y + x(1-x)]  ln(\frac{\ov{M_y^2}}{x(1-x) \Lambda^2})}
{[\ov{M_y^2}-x(1-x) \Lambda^2]}
 \right. \nonumber \\  &  & \left. - \frac{y x}{\ov{M_y^2}} + 
 \frac{y^2 (3 x-1)}{(1-x)\ov{M_y^2}} - \frac{x(1-x)}{3\ov{M_x^2}}
\right\}   \nonumber \\
&  & 
 +  3 m^2 \int_0^1 dx \int_0^{1-x} dy \left\{ 
 \frac{[(-3/2+2 x)y + x(1-x)]  ln(\frac{\ov{M_y^2}}{x(1-x) \Lambda^2})}
{[\ov{M_y^2}-x(1-x) \Lambda^2]}
 \right. \nonumber \\  &  & \left. - \frac{y(x+3y)}{\ov{M_y^2}} -
 \frac{3y^2 (1-y)}{x(1-x)\ov{M_y^2}}
 + \frac{2 y^3}{(1-x)^2 \ov{M_y^2}} -
\frac{x(1-x)}{3\ov{M_x^2}} \right\} \; \; , 
\label{ahat}
\eea
includes the unphysical Higgs contribution $\sim 3 m^2$.
Since we have omitted any current mass for the $s$- and $d$-quark
and expanded to third order in the external momentum $k$, one has to 
introduce a lower cut-off scale $\Lambda$. In order to  match  
 the low-energy  description in terms of the $\chi QM$, its magnitude
 will  be taken as  the chiral symmetry breaking scale
 $\Lambda_{\chi}$ = 0.83 GeV.  

 It should be noted that in the previous
calculation \cite{SPj}, only the case $m_q^2 \ll M_W^2$ was considered.
In this case only the $B_0$ term contributes to order $G_F$, while
 contributions
from the other $B_i$'s are suppressed by $m_q^2/M_W^2$ with respect to the
$B_0$ contribution. Including the heavy top quark, all contributions have
to be taken into account.

 Combining the equations (\ref{SP2})
and (\ref{ahat}), we find the self-penguin contribution to the quantity
$A$ in (\ref{LdsR0}):
\be
A_{SP}  \; = \; \frac{G_F}{\sqrt{2}} \, \frac{h}{24\pi^2} \,
             \frac{\alpha_s}{\pi}  
\left[ \lambda_u( F_u -F_c) -  \lambda_t (F_c - F_t) \right] \; ,
\label{selfpe}
\ee
where 
$h_c = Tr(t^a t^a)/N_c = (N_c^2 -1)/(2 N_c)$ is a colour factor, and
$N_c$ is the number of colours.
Eqs. (\ref{SP1})-(\ref{ahat}) is the contribution from the diagram in
Fig.~1 (including the unphysical Higgs contribution). Taking into account 
the counterpart where the gluon is
 absorbed by the $d$-quark, one gain a factor two which is included in
(\ref{selfpe}).
Numerically, we find for $m=m_c=1.4$ GeV and $\Lambda = 0.83$ GeV:
\be
F_c  \simeq 49.0 \; \; .
\label{Fc}
\ee
Similarly, for $m=m_t=180$ GeV and $\Lambda = 0.83$ GeV, we obtain 
\be
F_t \simeq  -10.7 \; \; .
\label{Ft}
\ee
For the $c$-quark case one obtains
the analytical result within the leading logarithmic approximation
\be
F_c  \; = \;
\frac{1}{2} [\ln(\frac{M_W^2}{\Lambda^2})]^2 \, - \,
\frac{1}{2} [\ln(\frac{m_c^2}{\Lambda^2})]^2  \, +  \, 
(\frac{5}{6} - \frac{4}{6})\ln(\frac{M_W^2}{m_c^2}) \, - \,
\frac{5}{6}\ln(\frac{m_c^2}{\Lambda^2})  \; .
\label{LLc}
\ee
The leading part of this result was obtained in \cite{SPj} for the 
charm quark case, while the general result (\ref{ahat}) is new.
In (\ref{LLc}) the term $\sim -4/6$ is coming from $-B_0 q^2 g^{\sigma \nu}$
in (\ref{Tsinu})and the rest from $B_0 q^{\sigma} q^{\nu}$.
As shown in \cite{QCDRGE,EKP}, leading logarithmic QCD corrections can be
taken into account by folding  the well known
 Wilson coefficients $C_{\pm}$ \cite{GaL,AlMa} into an integral over virtual
momenta $p$. Then our result takes the form
\be
\frac{\alpha_s}{\pi} F_c  \, \ra \tilde{F_c}\; = \;
 \int_{m_c^2}^{M_W^2} \frac{dp^2}{p^2} \rho(p^2)
 \left\{ \ln(\frac{p^2}{\Lambda^2}) \, + 
\, (\frac{5}{6}-\frac{4}{6}) \right\} 
\, - \, \frac{5}{6} \rho(m_c^2) \ln(\frac{m_c^2}{\Lambda^2}) \; ,
\label{cRGE}
\ee
where $\rho(p^2) = [C_+(p^2) + C_-(p^2)] \alpha_s(p^2)/(2\pi)$.
Because the operator in (\ref{LdsR0}) has zero anomalous dimension, there
 is no extra factor attached to $\rho(p^2)$.
The linear combination $(C_+ + C_-)/2$ is the colour octet combination
of the coefficients which occurs because the gluon within the
self-penguin is a colour octet. The $C_\pm$ 's are given by ratios of
$\alpha_s$ at different scales to a power determined by the
anomalous dimension of the corresponding four quark operators $Q_\pm$.
Thus the integral in (\ref{cRGE}) can be performed analytically, and the
result obtained in terms of $\alpha_s$ at the scales $\Lambda, m_c, m_b$,
and $M_W$ respectively.
Numerically, we find $\tilde{F_c} \simeq 3.25$. For the top quark 
contribution, we expect essentially no QCD corrections except for the running
 of $\alpha_s$, and we take
\be
 \tilde{F_t} \, = \, \frac{\alpha_s(m_t^2)}{\pi} F_t \simeq -0.37 \; ,
\label{tRGE}
\ee
 approximately the same as the pure EW  contribution. The total $A$
to be used in (\ref{LdsR0})  is $A_{SP} + A_{EW}$. (There are other
 contributions of order $\alpha_s G_F$, but with a very small
 numerical coefficient. Note also that $A_{EW}$ is dominated by loop momenta
of order $M_W$ where $C_{\pm} \simeq 1$, and there will be no further QCD corrections in the
leading logarithmic approximation).

\vspace{0.5cm}
\begin{Large}

3. The Chiral Quark Model

\end{Large}

\vspace{0.2cm}

In the Chiral Quark Model ($\chi QM$) \cite{ChiQM},
 chiral-symmetry breaking is 
taken into account by adding an extra term 
${\cal{L}}_{\chi}$ to ordinary QCD:
\begin{equation}
  {\cal{L}}_{\chi} = - M \, (\overline{q_L} \Sigma^\dagger q_R + 
\overline{q_R} \Sigma q_L) \; ,
\label{Lchi}
\end{equation}
where $\bar{q} = (\bar{u}, \bar{d}, \bar{s})$. The constant  $M$
 in (\ref{Lchi}) is interpreted as the
{\em constituent} quark mass, expected to be of order  200-300 MeV.
The quantity $\Sigma$ contains
 the Goldstone-octet fields $\pi_a$:
\begin{equation}
\Sigma = exp(i\sum_a \lambda_a \pi_a / f) \; \; ,
\label{sig}
\end{equation}
where the $\lambda_a$'s
are the Gell-Mann matrices, and $f  =  f_{\pi} = 93$ MeV is the pion
decay constant.
 The term ${\cal{L}}_{\chi}$ introduces a meson-quark coupling.
This means that the quarks can be integrated out and the coefficients
of the various terms in  the chiral Lagrangian are calculable \cite{ChiQM}.

The $\chi QM$ should be 
applied  for  momenta lower than 
the scale of chiral-symmetry breaking, 
$\Lambda_{\chi} = 2 \pi f_\pi \sqrt{6/N_c}$, where $N_c = 3$ is
the number of colours. Divergent integrals are regularized, for instance
with dimensional regularization \cite{BEF}.
An alternative is to use
some  ultraviolet cut-off  $\Lambda_U$ of the  order $\Lambda_{\chi}$.
 Because $f_{\pi}$, entering the meson-
quark coupling $\sim M \gamma_5/f$, is also given by a
quark loop diagram for $\pi \ra W$(virtual), one has:
\begin{equation}
f^{(0)}_{\pi} \, = \, \frac{N_c M^2}{4 \pi^2 f} 
[ \; \hat{f}_{\pi}
\;  + \frac{\pi^2}{6 N_c M^4}
<\frac{\alpha_s}{\pi} G^2> + \cdots ] \; ,
\label{fpi}
\end{equation}
where  $<\frac{\alpha_s}{\pi} G^2>$
 is the two gluon condensate, and the dots indicate higher gluon
condensates. In the end both $f$ and $f^{(0)}_{\pi}$ will, in the
limit $m_{u,d} \ra 0$, be identified by $f_{\pi}$, but at
intermediate stages one might need to distinguish them for
technical reasons.
The dimensionless quantity $\hat{f}_{\pi}$ has the leading behaviour 
$\sim \ln(\Lambda_U^2/M^2)$ in a cut-off type
regularization and $\sim \Gamma(\epsilon) \sim  1/\epsilon$ 
 within dimensional regularization
($D = 4 - 2 \epsilon$ being the dimension of space).

The quark condensate can be written as
\begin{equation}
<\bar{q} q> \,  \: = \:  \frac{N_c M}{4 \pi^2} C_q
 - \frac{1}{12 M} <\frac{\alpha_s}{\pi} G^2> + \cdots  \ ,
\label{qcond}
\end{equation}
 The  quantity
$C_q$ depends on  the regularization prescription.
Using a cut-off $\Lambda$, one finds $C_q = - \Lambda^2  + 
M^2 \ln (\Lambda_U^2/M^2) + \cdots$ , and
within dimensional regularization, $C_q = - M^2 \Gamma(-1 + \epsilon) +
 \cdots$.


\vspace{0.7cm}

\begin{Large}

4. The  $K \rightarrow 2 \pi$ amplitude from $s \rightarrow d$. 

\end{Large}


\begin{figure}
\begin{center}
\mbox{\epsfig{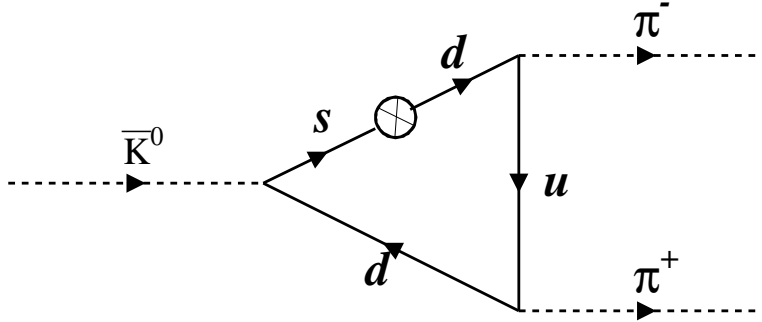}}
\end{center}
\label{k2pisdselfp}
\caption{Typical diagram for non-diagonal $s \ra d$ self-energy
  induced $K \ra 2\pi$ transition.}
\end{figure}

\vspace{0,2cm}
Within the $\chi QM$, the  $K \rightarrow 2 \pi$ amplitude 
obtained from the $s \rightarrow d$ transition in (\ref{LdsR0})
is determined by diagrams like the one in Fig.~2.
We find  the following result:
\be
 {\cal M}(\ov{K^0} \ra \pi^+ \pi^-)_\Sigma =
 \frac{ \sqrt{2}}{4f_{\pi}^3} (m_K^2 - m_\pi^2)
A M^2  \left[ \frac{<\bar{q} q>}{M} \, + \, 2 f_\pi^2 
\left(1 -  \frac{3 M^2}{2 \Lambda_\chi^2} \right) \right] \, ,
\label{SPA}
\ee
where $A$ is the quantity in  (\ref{LdsR0}).
For completeness, we have also calculated the virtual $K \ra vacuum$
 (which is zero) and
$K \ra \pi$ transition amplitudes  in addition to the physical
$K \ra 2\pi$ amplitude. These calculations show that 
${\cal L}_{ds}^R$ in (\ref{LdsR0})
gives an octet contribution that can be simply added to the $Q_6$
contribution up to ${\cal O}(p^2)$. To order ${\cal O}(p^4)$ we expect to
obtain different contributions from $Q_6$ and 
${\cal L}_{ds}^R$   due to the extra
derivatives in ${\cal L}_{ds}^R$. 

 We want to compare the result (\ref{SPA}) with the corresponding
  amplitude for the penguin operator $Q_6$ :
\begin{equation}
{\cal M}(\overline{K^0} \rightarrow \pi^+ \pi^-)_{Q_6} \; = \; 
 \frac{ 2 \sqrt{2}}{f_\pi} (m_K^2-m_{\pi}^2) \, C_6 \,  
\frac{ < \bar{q} q> }{M} \left[ 1- 6 \frac{M^2}{\Lambda_\chi^2} \right] \; .
\label{Q6A}
\end{equation}
 We observe that the
ratio between the amplitudes in (\ref{SPA}) and (\ref{Q6A}) goes to
 zero in the limit where the constituent quark mass 
$M \rightarrow 0$, as expected.

Numerically, the ratio between the self-penguin and the $Q_6$ 
contribution is less than 5 $\%$  in the CP-conserving
case. 
For $\epsilon'/\epsilon$, the situation is different
 because  there is a large cancellation between the gluonic and electroweak
 penguin contributions\cite{BuFR,NLO,BEF}. In the operator
 language, it is a large cancellation between the $Q_6$ and the $Q_8$
contributions.
We have compared 
the $s \ra d$ self-energy contribution with the standard $Q_6$ contribution
in the CP-violating case. Using, in the notation of \cite{NLO},  
$C_i = - (G_F/\sqrt{2})(\lambda_u z_i - \lambda_t y_i)$, with  
$y_6 = - 0.137$ at the renormalization scale $\mu = \Lambda = \Lambda_\chi$;
 $M \simeq$ 220 MeV and 
$<\bar{q} q>^{1/3} \simeq -220$ MeV \cite{BEF}, 
 we find that in the CP-violating case,  the $s \ra d$ contribution
is  10 to 15 $\%$ of the $Q_6$ contribution and has the same sign.
 This means that after the partial cancellation between 
 the $Q_6$ and the $Q_8$ contributions has been taken into account,
the $s \ra d$ contribution is of the order
$\epsilon'/\epsilon$ itself.

\vspace{0.5cm}

\begin{Large}

5. Discussion

\end{Large}

\vspace{0.2cm}

We have considered the $s \ra d$ self-energy  contribution to
 $K \rightarrow 2 \pi$
within the chiral quark model. Within this model matrix elements between the light
mesons ($\pi, K$) of quark operators may be calculated in terms
of quark loops.
 Therefore this
model  provides a means to calculate effects of  off-shell quarks.

 In the literature, terms which vanish by using the perturbative equations 
of motion are omitted from the analysis from the very
beginning\cite{AlMa,CoDDPoS,OP}.  But to neglect such effects
 would, within QED,
be analogous to discarding Lamb-shift. Moreover, such effects are larger
in QCD where the bound state effects are stronger.
Still, also in strong interactions such effects are relatively small,
 being proportional to $M^2/\Lambda_\chi^2 \simeq 0.07$ and vanishing in
the free quark limit $M \ra 0$.
 They  might, however,  become non-negligible in cases where there are
 cancellations between the potentially largest terms, as shown in this paper
for  $\epsilon'/\epsilon$ , and for $K \ra 2 \gamma$ \cite{offsh} where
the potentially large pole contributions cancel when the 
Gell-Mann-Okubo mass formula is used.

 Numerically, we have found a small  effect for the CP-conserving
 $K\rightarrow 2\pi$ amplitude. However, for the CP-violating
 $K\rightarrow 2\pi$ amplitude and thereby for   $\epsilon'/\epsilon$
 the effect is non-negligible, being of the order  $\epsilon'/\epsilon$
itself,  because of the strong cancellation between the $Q_6$ and $Q_8$
contributions. It would  of course have been an advantage if
${\cal L}_{ds}^R$ were  included in the basis of a complete
next to leading QCD corrections \cite{NLO}, but this would have required a
three loop calculation. 
The self-penguin contribution drives $\epsilon'/\epsilon$ more to the
positive side. However, due to the uncertain value of the quark condensate
it is still difficult to make a very precise prediction of
  $\epsilon'/\epsilon$ within the chiral quark model\cite{BEF}.


\end{document}